\documentclass[utf8]{FrontiersinHarvard} % for articles in journals using the Harvard Referencing Style (Author-Date), for Frontiers Reference Styles by Journal: https://zendesk.frontiersin.org/hc/en-us/articles/360017860337-Frontiers-Reference-Styles-by-Journal
\usepackage{url,hyperref,lineno,microtype,subcaption}
\usepackage[onehalfspacing]{setspace}
\usepackage{xcolor}

\def\keyFont{\fontsize{8}{11}\helveticabold }
\def\firstAuthorLast{Yadav {et~al.}} %use et al only if is more than 1 author
\def\Authors{Nitin Yadav\,$^{1,*}$ and Apanba Khuman\,$^{1,*}$}
\def\Address{$^{1}$Department of Physics, Indian Institute of Technology Delhi, Delhi 110016, India.}
\def\corrAuthor{Nitin Yadav}

\def\corrEmail{nitnyadv@gmail.com}

\begin{document}
\onecolumn
\firstpage{1}

\title[Vortex Influence on Slow Magnetoacoustic Waves]{Solar Vortices: Catalysts of Magnetoacoustic Wave Dissipation and Atmospheric Heating} 

\author[\firstAuthorLast ]{\Authors} %This field will be automatically populated
\address{} %This field will be automatically populated
\correspondance{} %This field will be automatically populated

\extraAuth{}% If there are more than 1 corresponding author, comment this line and uncomment the next one.
%\extraAuth{corresponding Author2 \\ Laboratory X2, Institute X2, Department X2, Organization X2, Street X2, City X2 , State XX2 (only USA, Canada and Australia), Zip Code2, X2 Country X2, email2@uni2.edu}

\maketitle

\begin{abstract}

%%% Leave the Abstract empty if your article does not require one, please see the Summary Table for full details.
\section{}
The propagation and dissipation of magnetohydrodynamic waves play a key role in transporting energy from the solar photosphere to the chromosphere. Using high-resolution three-dimensional radiative MHD simulations, we investigate the evolution of slow magnetoacoustic waves along magnetic field lines and examine the influence of photospheric vortex flows on wave dynamics and heating. Field-line tracking reveals upward-propagating slow-mode waves that amplify in the stratified atmosphere and steepen into shocks in the chromosphere, producing recurrent plasma surges with characteristic chromospheric shock signatures. 
% \textcolor{blue}{Shock locations and thickness are quantified using gradients in density perturbations}.
Vortex regions are identified using the swirling strength diagnostic with height-dependent Gaussian smoothing to capture expanding vortex structures. A comparison between vortex and non-vortex field lines shows systematically enhanced temperature in vortex regions.Furthermore, a comparison of shock formation height between vortex and non-vortex regions reveals no systematic difference, indicating that rotational flows do not significantly alter the height at which shocks form. However, supersonic upflows at vortex locations exhibit somewhat higher parallel velocities compared to non-vortex regions, suggesting that vortex-driven motions may amplify the velocity of propagating shocks.
These results indicate that vortex-driven motions contribute to increased shock dissipation and modify the thermal structure of the lower solar atmosphere, highlighting the coupled role of slow-mode shocks and vortex flows in chromospheric energy transport.

\tiny
 \keyFont{ \section{Keywords:}  Magnetohydrodynamics (MHD), Slow magnetoacoustic waves, Photospheric vortices, Shock formation, Energy transport, Solar chromosphere, Numerical simulations} %All article types: you may provide up to 8 keywords; at least 5 are mandatory.
\end{abstract}

\section{Introduction}
Understanding the mechanisms responsible for transporting energy and momentum from the solar photosphere to the overlying chromosphere and corona remains a long-standing challenge in solar physics. Magnetohydrodynamic (MHD) waves are widely considered to play a fundamental role in this process, particularly in magnetised regions where the magnetic field governs plasma motions and provides preferential channels for wave propagation \citep{Alfven1947, NakariakovVerwichte2005, Jess2023,Kumar2024, Nakariakov2024}. Theoretical studies predict, and observations confirm, the presence of a rich spectrum of MHD wave modes in the solar atmosphere, including Alfvén, fast magnetoacoustic, and slow magnetoacoustic waves \citep{Roberts2000, DeMoortelNakariakov2012, grant2022, Liu2023}. These waves are primarily excited by convective motions in the photosphere—such as granular flows, global oscillations, and magnetic footpoint perturbations—and can propagate upwards along magnetic field lines into higher atmospheric layers. Recent advances in high-resolution ground- and space-based instrumentation have enabled the detection of MHD waves across a wide range of spatial and temporal scales, demonstrating their ubiquity in magnetic elements, sunspots, spicules, fibrils, and coronal loops \citep{Tomczyk2007, Jess2009, Morton2012, Antolin2014, Morton2023, Jess2023, Zhou2024}.

Within the gravitationally stratified solar atmosphere, MHD waves are expected to play an important role in the transport and redistribution of energy across atmospheric layers. As these waves propagate through regions characterised by strong gradients in density, temperature, and magnetic field strength, their behaviour is significantly modified by processes such as reflection, refraction, mode conversion, and dissipation \citep{Bogdan2003, Cally2007, Khomenko2022,Felipe2022, Snow2022}. Compressive wave modes, in particular, are susceptible to nonlinear steepening as they propagate upwards, leading to the formation of shocks in the chromosphere. Such shock-driven dynamics have been shown to produce intermittent heating and enhanced chromospheric variability \citep{CarlssonStein1997, Felipe2010, Srivastava2010,Bose2022}. Consequently, a detailed understanding of how local magnetic topology and plasma flows influence MHD wave propagation, mode coupling, and dissipation is essential for reliably assessing the role of waves in atmospheric heating and in coupling the photosphere to the upper solar atmosphere.

Among the MHD wave modes present in the solar atmosphere, slow magnetoacoustic waves are of particular relevance in the lower atmospheric layers, where plasma motions are strongly guided by the magnetic field. These compressive waves propagate predominantly along magnetic field lines and are frequently observed in photospheric and chromospheric magnetic structures, including pores, sunspots, network elements, and slender magnetic flux tubes \citep{Bogdan2003, KhomenkoCollados2015,Jafarzadeh2022,Chae2023}. As slow magnetoacoustic waves propagate upward through the gravitationally stratified atmosphere, their amplitudes increase, often leading to nonlinear steepening and the formation of shocks. Both observational and numerical studies have shown that such shock formation is a common feature of slow-mode wave propagation and can contribute significantly to chromospheric heating and dynamic phenomena, particularly in regions of enhanced magnetic flux \citep{CarlssonStein1997, Centeno2006, Felipe2011, Bose2022, Chae2023}.

In parallel with the growing body of work on MHD waves, high-resolution observations and numerical simulations have revealed that vortex flows are a ubiquitous feature of the solar photosphere, particularly within intergranular lanes where converging horizontal motions and associated downdrafts give rise to localized rotational flows \citep{Bonet2008, VargasDominguez2011, Tziotziou2020, Dakanalis2022, DiazCastillo2024}. These photospheric vortices are capable of twisting and braiding magnetic field lines, thereby injecting magnetic energy and helicity into the overlying atmosphere \citep{Shelyag2011, Wedemeyer2013,Shetye2022}. Recent studies using realistic magnetohydrodynamic simulations have further characterized the formation, evolution, and energetics of vortex structures, highlighting their potential role in mediating wave excitation and small-scale dynamic phenomena \citep{Yadav2020, Yadav2021vortices, YadavKannan2024, Shen2025}. As a consequence, such vortical motions have been proposed as efficient drivers of MHD wave activity, including torsional Alfvén waves and magnetoacoustic perturbations, and as contributors to localized atmospheric heating through enhanced wave energy flux and dissipation \citep{Fedun2011, Moll2012, YadavKannan2024}. Despite mounting observational and numerical evidence for their prevalence and dynamical significance, the extent to which photospheric vortices influence the excitation, propagation, and evolution of waves in the lower solar atmosphere remains insufficiently constrained.

The interaction between photospheric vortex flows and slow magnetoacoustic waves represents a particularly compelling aspect of wave–flow coupling in the lower solar atmosphere. Vortical motions can locally modify the magnetic field configuration as well as the plasma density and pressure stratification, thereby affecting wave excitation, propagation, mode conversion, and the onset of nonlinear steepening and shock formation \citep{Bogdan2003, Fedun2011, Yadav2021, Kesri2024}. Furthermore, vortex-driven perturbations of magnetic field lines may act as localized wave sources or alter the effective waveguiding properties of magnetic structures, potentially enhancing or suppressing the upward transmission of slow magnetoacoustic waves into the chromosphere \citep{Shelyag2011, Keys2013, KhomenkoCollados2015, Liu2023}. Despite these theoretical expectations and supporting numerical results, observational constraints remain limited, and systematic studies quantifying how photospheric vortices influence slow-mode wave properties and their evolution into shocks are still lacking. In particular, the Daniel K. Inouye Solar Telescope (DKIST) is ideally suited to detecting photospheric vortices and their signatures in the chromosphere, as well as tracking the propagation of slow magnetoacoustic waves along magnetic flux tubes \citep{Rimmele2020,Rast2021}. High-resolution spectropolarimetric and imaging observations can reveal how localized rotational flows twist magnetic field lines and excite wave modes, providing observational constraints on mechanisms that have so far been explored primarily in numerical simulations.

Previous investigations have shown that slow magnetoacoustic waves in plage and network regions can transport significant energy capable of heating the chromosphere \citep{Yadav2021}. These studies employed field-line tracking combined with power analysis to follow the evolution of waves along magnetic flux tubes. This approach allows wave properties to be measured along the actual path of energy propagation, rather than across fixed horizontal or vertical slices, providing a more accurate representation of wave amplitudes, phase speeds, and nonlinear steepening. By capturing the full three-dimensional geometry of magnetic structures, field-line tracking also enables a precise determination of where shocks form and how energy is transported through the lower solar atmosphere. Building on these findings, we apply similar field-line tracking to both dynamical and thermodynamic variables, enabling a more comprehensive analysis of wave evolution and shock formation along magnetic field lines while maintaining methodological consistency with previous work.

In this paper, we examine how the presence of photospheric vortex flows influences the propagation of slow magnetoacoustic waves and the formation of shocks in the lower solar atmosphere. By comparing magnetic regions exhibiting clear vortex signatures with regions lacking such rotational motions, we quantify differences in temperature, viscous and Joule heating rates. Through this analysis, we aim to clarify the role of photospheric vortices in mediating the coupling between photospheric dynamics and chromospheric wave activity, thereby improving our understanding of small-scale energy transport processes in the solar atmosphere. By establishing how the presence of vortices modifies thermal and dynamical evolution of the atmosphere, our results provide predictive diagnostics that can be tested with forthcoming DKIST observations, bridging the gap between numerical modeling and high-resolution solar measurements.

The remainder of this paper is structured as follows. In Section \ref{methods}, we describe the simulation setup and the field-line tracking methodology used to analyse slow magnetoacoustic wave propagation and shock formation in the lower solar atmosphere. Section \ref{results} presents the results, including wave propagation, shock occurrence, swirling strength and vortex influence of the lower atmosphere. Finally, Section \ref{summary} summarises our key findings, highlights their significance in the context of vortex–wave interactions, and outlines directions for future observational and numerical investigations.

\section{Data and Methodology}\label{methods}
To investigate the propagation of slow magnetoacoustic waves and the influence of photospheric vortices, we employ the three-dimensional radiative MHD code MURaM \citep{Vogler2005, Rempel2014, Rempel2017}. The code solves the compressible MHD equations for a partially ionized plasma in a Cartesian box, including radiative transfer and realistic equations of state. This approach allows self-consistent modeling of turbulent convection in the upper solar convective zone and the lower atmosphere, naturally exciting MHD waves without the need for imposed drivers. For this study, we focus exclusively on a strong plage region with a mean vertical magnetic field of 200 G. The simulation domain spans 12 Mm × 12 Mm horizontally and 4 Mm vertically, resolved with 10 km grid spacing in all directions. 
The lower boundary is located 1.5~Mm below the mean solar surface and is treated as open to permit convective inflows. The top boundary, positioned 2.5~Mm above the mean surface, is implemented as an open boundary to allow for outflows, with a potential-field extrapolation applied to the magnetic field. Periodic boundary conditions are imposed in the horizontal directions. The domain spans the upper convection zone, photosphere, and chromosphere.
The simulations are evolved over several convective turnover times to ensure that a quasi-stationary state is attained before extracting data for further analysis.
% The simulations are initialized with a purely hydrodynamic run for 2 h to achieve a relaxed convective state, after which the 200 G vertical magnetic field is introduced and the simulation is run for an additional 1.2 h until a statistically stationary state is reached. 
From this evolved state, we extract a 20 min time series at 1 s cadence from a 4 Mm × 4 Mm × 3 Mm sub-region extending from 0.5 Mm below the mean surface to 2.5 Mm above it. This sub-region captures the dynamics of a strong magnetic concentration while maintaining high temporal and spatial resolution, sufficient to resolve slow-mode waves and their nonlinear steepening into shocks.

To study wave propagation along magnetic structures, we employ a field-line tracking methodology \citep{Yadav2021}. Seed points are placed within the magnetically concentrated region at the photospheric surface, and magnetic field lines are traced in 3D at each time step using the vector magnetic field. The seed points are advected by the local velocity field at z = 0 Mm to capture the motion of plasma parcels, after which the field lines are re-traced from their updated positions. This procedure is repeated throughout the 20 min time sequence, allowing us to follow the slow magnetoacoustic waves as they propagate along dynamically evolving magnetic structures. In addition to the parallel velocity, we extend this analysis to thermodynamic variables, including density, pressure, and temperature, as well as the velocity divergence, thereby providing a more detailed view of the compressive wave dynamics along field lines. 
% \textcolor{blue}{Shock formation is identified through the development of steep gradients in the field-aligned profiles, and we quantify the shock properties by measuring the shock height and thickness from the maximum vertical gradient of the density perturbation profile}

In the present study, we further analyse the influence of photospheric vortices on the slow magnetoacoustic wave propagation by calculating the swirling strength across different layers of the simulation domain. To detect coherent vortex structures while minimizing noise from small-scale turbulent motions, the velocity field is first smoothed using a Gaussian convolution with height dependent smoothing width. This approach allows us to detect expanding vortex structures spanning multiple heights in the lower solar atmosphere. To examine the thermodynamic response of the plasma, we further quantify temperature, viscous heating and resistive heating by taking average over field lines over time from both vortex and non-vortex regions. This allows us to compare the thermal and dissipative properties associated with these regions.
\section{Results and Discussion}\label{results}
\begin{figure}[h!]
\begin{center}
\includegraphics[width=0.8\textwidth]{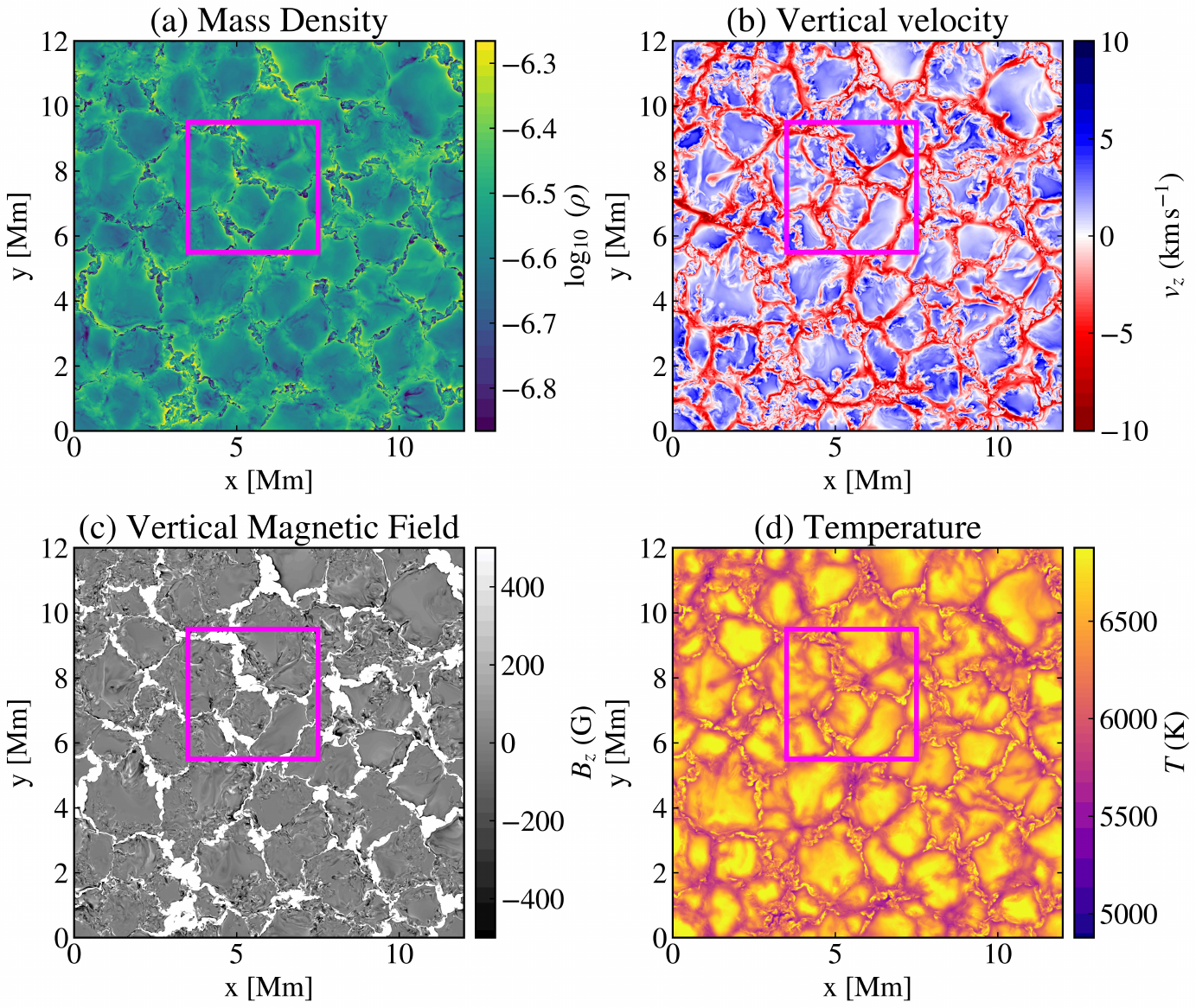}
\end{center}
\caption{Spatial distribution of physical parameters at the $\tau_{500,\mathrm{nm}} = 1$ (continuum formation) layer over the full simulation domain. Panels show (a) mass density, (b) vertical velocity, (c) vertical magnetic field strength, and (d) temperature.}\label{fig:1}
\end{figure}

Figure \ref{fig:1} shows the spatial distribution of key physical quantities at the 
$\tau_{500nm}=1$ layer (continuum formation height) over the full horizontal extent of the simulation domain. Panel (a) displays the mass density, panel (b) the vertical (z) component of the velocity, panel (c) the vertical component of the magnetic field, and panel (d) the temperature. The characteristic granular pattern is clearly visible, with hot, low-density upflows associated with granules surrounded by cooler, denser downflows in the intergranular lanes. Magnetic fields are advected by the convective plasma motions and become concentrated within the intergranular lanes, where the vertical magnetic field strength reaches values of up to $\sim$ 1.5 kG. The temperature distribution largely follows the convective pattern, with higher temperatures in granules and relatively lower temperatures in intergranular downflows. However, localized regions of enhanced temperature are also evident within strongly magnetized intergranular lanes. These temperature enhancements arise from the Wilson depression, whereby the optical depth unity surface is displaced downward in strong magnetic concentrations, exposing deeper and hotter layers of the solar atmosphere.

\begin{figure}[h!]
\begin{center}
\includegraphics[width=0.8\textwidth]{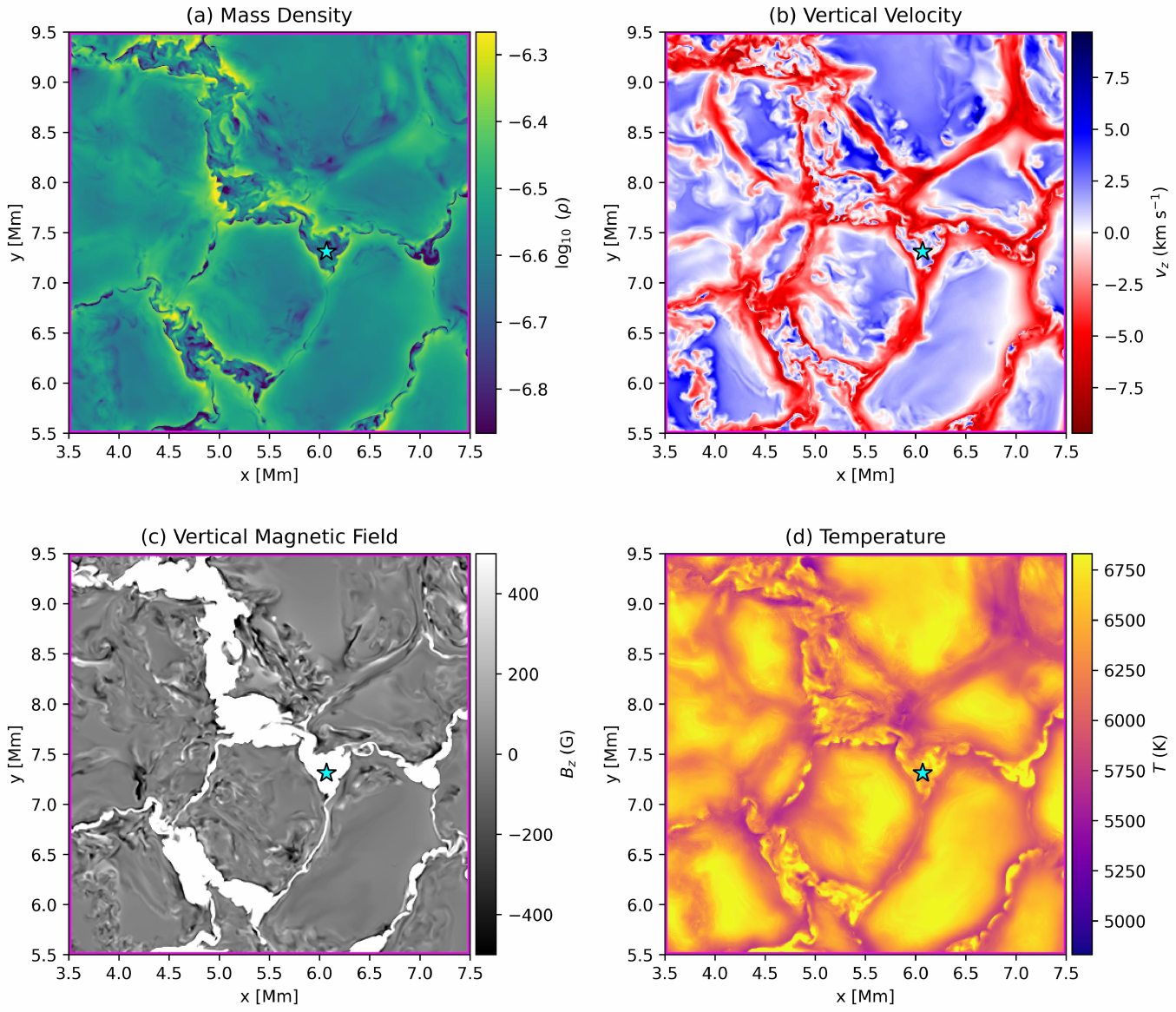}
\end{center}
\caption{Enlarged view of the 4 Mm × 4 Mm sub-region (indicated by the magenta square in Figure 1) selected for detailed time-series analysis. The region includes strong magnetic concentrations embedded within convective flows, providing a representative environment for examining vortex dynamics and the propagation of slow magnetoacoustic waves. The overlaid cyan asterisk indicate the location of seed point from which the associated magnetic field line is traced in both space and time for further examination.}\label{fig:2}
\end{figure}

The magenta square overlaid in Figure \ref{fig:1} marks a sub-region of size 4 Mm × 4 Mm, selected for detailed time-series analysis of wave propagation and vortex activity. This region encompasses strong magnetic concentrations as well as surrounding convective flows, providing a suitable environment for investigating the interaction between photospheric dynamics and slow magnetoacoustic waves. A magnified view of this sub-region is presented in Figure \ref{fig:2}, where the fine-scale magnetic and flow structures relevant to the subsequent analysis are shown in greater detail. A cyan asterisk is overlaid to indicate the location of seed point placed at z = 0 Mm layer from which the associated magnetic field line is traced in both space and time for further examination. 
A movie accompanying Figure \ref{fig:2} is also available in the online version, illustrating the evolution of these fields with height, along with the corresponding motion of the traced magnetic field line.

\begin{figure}[h!]
\begin{center}
\includegraphics[width=0.7\textwidth]{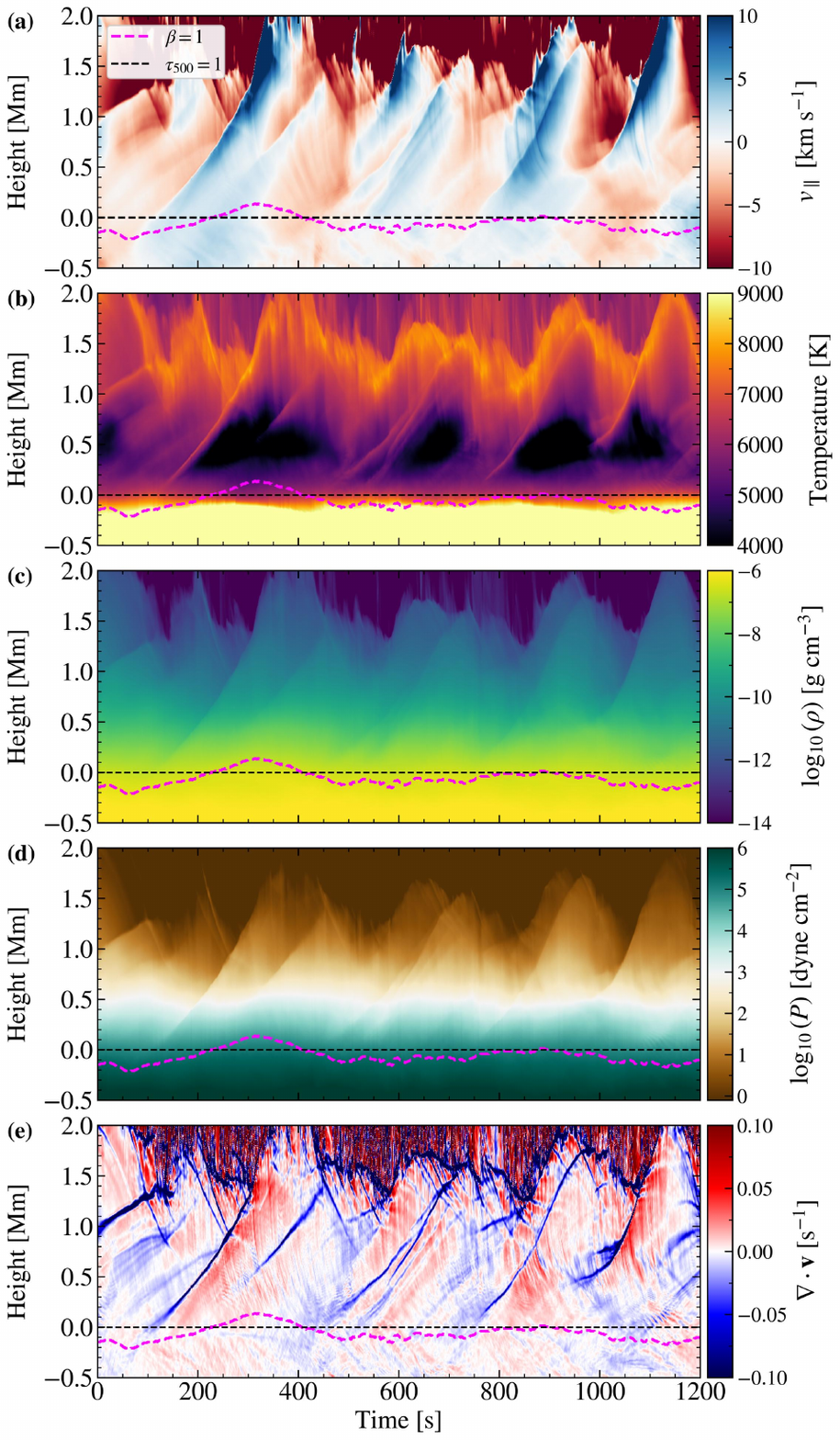}
\end{center}
\caption{Height–time diagrams of plasma parameters along a selected magnetic field line, shown in a frame shifted relative to the $\tau_{500}=1$ layer (dashed line). Panels show (a) field-aligned velocity $v_\parallel$, (b) temperature, (c) $\log_{10}(\rho)$, (d) $\log_{10}(P)$, and (e) velocity divergence $\nabla \cdot \mathbf{v}$. Upward-propagating compressive disturbances and their amplification into shock-like features in the upper layers are evident.The $\beta = 1$ layer is also indicated by a dashed magenta curve.}\label{fig:3}
\end{figure}

Figure \ref{fig:3} illustrates the temporal evolution of key thermodynamic and dynamic quantities along this representative magnetic field.
The maps are shown in a height coordinate shifted with respect to the $\tau_{500nm}=1$ surface, such that $z' = z - z_{\tau=1}$ and $z' = 0$ denotes the optical depth unity layer. This transformation removes fluctuations associated with the corrugation of the photosphere and allows us to track wave propagation relative to a physically meaningful reference layer.
The dashed magenta line indicates the plasma $\beta = 1$ layer. As seen in the figure, the majority of the region of interest, viz., the photosphere and chromosphere sampled in our analysis lies predominantly in the $\beta < 1$ regime. In a low-$\beta$ environment, slow magnetoacoustic waves propagate primarily along magnetic field lines; therefore, we use the parallel velocity component as a proxy of slow wave propagation. Although wave mode conversion can occur near the $\beta = 1$ layer, the regions analyzed in this study are largely situated above this equipartition layer. Consequently, the velocity component parallel to the background field is mostly associated with slow magnetoacoustic wave.
The upper portion of the domain is now shown, as results in this region are less reliable owing to their proximity to the top boundary, which may introduce non-physical boundary effects.

This figure reveals recurring upward surges of dense material with an approximate periodicity of five minutes, each exhibiting a characteristic parabolic trajectory reaching heights of 1--2~Mm before the plasma falls back under gravity. This periodicity is consistent with solar p-mode oscillations, which originate as nearly sinusoidal acoustic waves in the lower atmosphere and steepen into slow magnetoacoustic shocks as they propagate upward into regions of decreasing density. The repeated density enhancements visible in the figure trace the passage of these shocks, each imparting an upward impulse to the cooler, denser chromospheric plasma along the magnetic field.
The resulting plasma motions follow quasi-ballistic trajectories, indicating that the dynamics between successive shock passages are largely governed by solar gravity, with the magnetic field primarily guiding the flow. While these shock-driven, field-aligned motions are qualitatively similar to those associated with Type~I spicules reported in previous studies \citep{DePontieu2004,Skirvin2024,Srivastava2025}, the present model is restricted to the photosphere and chromosphere and does not include a transition region or coronal layers. Therefore, we interpret these dynamics more conservatively as chromospheric responses to p-mode driven slow magnetoacoustic shocks, without making a direct association with fully developed spicules.

% This figure reveals recurring upward surges of dense material with an approximate periodicity of five minutes, each exhibiting a characteristic parabolic trajectory reaching heights of 1–2 Mm before the plasma falls back under gravity. This periodicity is consistent with the well-known solar p-mode oscillations, which originate as nearly sinusoidal acoustic waves in the lower atmosphere and steepen into slow-mode shocks as they propagate upward into regions of progressively lower density. The repeated density enhancements visible in the figure thus trace the passage of these steepened slow shocks, each imparting a ballistic upward impulse to the cooler, denser chromospheric plasma along the field line. The parabolic form of the trajectories confirms that the plasma motion between successive shock passages is dominated by solar gravity, with the field line serving primarily as a guide rather than exerting significant additional force during the ballistic phase.
The field-aligned velocity ($v_\parallel$) clearly reveals upward-propagating disturbances originating below or near the $\tau_{500nm}=1$ level. In the lower photospheric layers ($z' \lesssim 0.3$ Mm), the perturbations exhibit relatively small amplitudes and propagate with nearly constant phase speed, consistent with linear slow magnetoacoustic waves traveling along the magnetic field. As the waves propagate upward into regions of decreasing density, their amplitudes increase significantly due to atmospheric stratification. The wave fronts progressively steepen, and in the mid-to-upper chromosphere ($z' \gtrsim 1.3$–2 Mm), they develop sharp gradients indicative of shock formation.
The thermodynamic variables, such as, temperature, mass density, and gas pressure, etc. display coherent, in-phase fluctuations with the velocity signal, confirming the compressive nature of the disturbances and their slow magnetoacoustic wave nature. Temperature enhancements coincide with compression phases, while density and pressure show corresponding increases, followed by rarefaction signatures. The logarithmic representation of density and pressure highlights the rapid decrease of background plasma parameters with height and emphasizes the relative amplification of perturbations in the upper layers.
The velocity divergence further substantiates the interpretation of these disturbances as compressive waves. Alternating bands of negative and positive $\nabla \cdot \mathbf{v}$ correspond to compression and rarefaction, respectively. In the chromospheric layers, the compression regions become spatially narrower and more intense, consistent with nonlinear steepening and shock development. The close spatial and temporal correspondence between strong negative divergence, enhanced temperature, and sharp velocity gradients supports the presence of upward-propagating acoustic shocks.
Overall, the combined signatures across all panels demonstrate the transformation of initially linear slow magnetoacoustic waves in the lower atmosphere into nonlinear, shock-like disturbances in the chromosphere. This evolution is primarily driven by density stratification and the associated amplification of wave amplitudes with height, highlighting the role of field-guided compressive waves in transporting energy from the photosphere into higher atmospheric layers.
% This figure illustrates results for a representative magnetic field line, with qualitatively similar maps obtained for other field lines. Additional examples are provided in the supplementary material. It is important to note that the typical lifetimes of chromospheric vortices are only a few minutes, whereas the duration over which wave propagation is tracked along a given field line in this study is significantly longer (on the order of $\sim$20 minutes). Consequently, individual field lines do not remain confined to either vortex or non-vortex regions, but instead traverse multiple such regions over time.
% We emphasize that the results presented are representative rather than specific to a single field line. We have performed similar analyses using multiple seed points distributed across the domain and find that the qualitative behavior, such as wave propagation, shock formation, and associated plasma response, remains consistent across these cases. Height time maps for a few additional field lines are provided in the supplementary material.

This figure illustrates results for a representative magnetic field line traced from a magnetically strong region at z = 0 Mm, with its starting location marked in Figure \ref{fig:2}. This field line is selected purely based on magnetic field strength. Although the field line is selected independently of vortex activity, it intermittently passes through vortex regions over the course of the full time series under evaluation. To demonstrate that the observed behaviour is not specific to a single field line, we performed the same analysis for various additional field lines, also traced from magnetically strong regions at z = 0 Mm without any vortex constraint. Height–time maps for three additional field lines are provided in the supplementary material showing qualitatively similar results in terms of wave propagation, shock formation, and associated plasma response. It is important to note that the typical lifetimes of chromospheric vortices are only a few minutes, whereas the duration over which wave propagation is tracked along a given field line in this study is significantly longer (on the order of $\sim$20 minutes). Consequently, individual field lines do not remain confined to either vortex or non-vortex regions, but instead traverse multiple such regions over time.

\begin{figure}[h!]
\begin{center}
\includegraphics[width=0.9\textwidth]{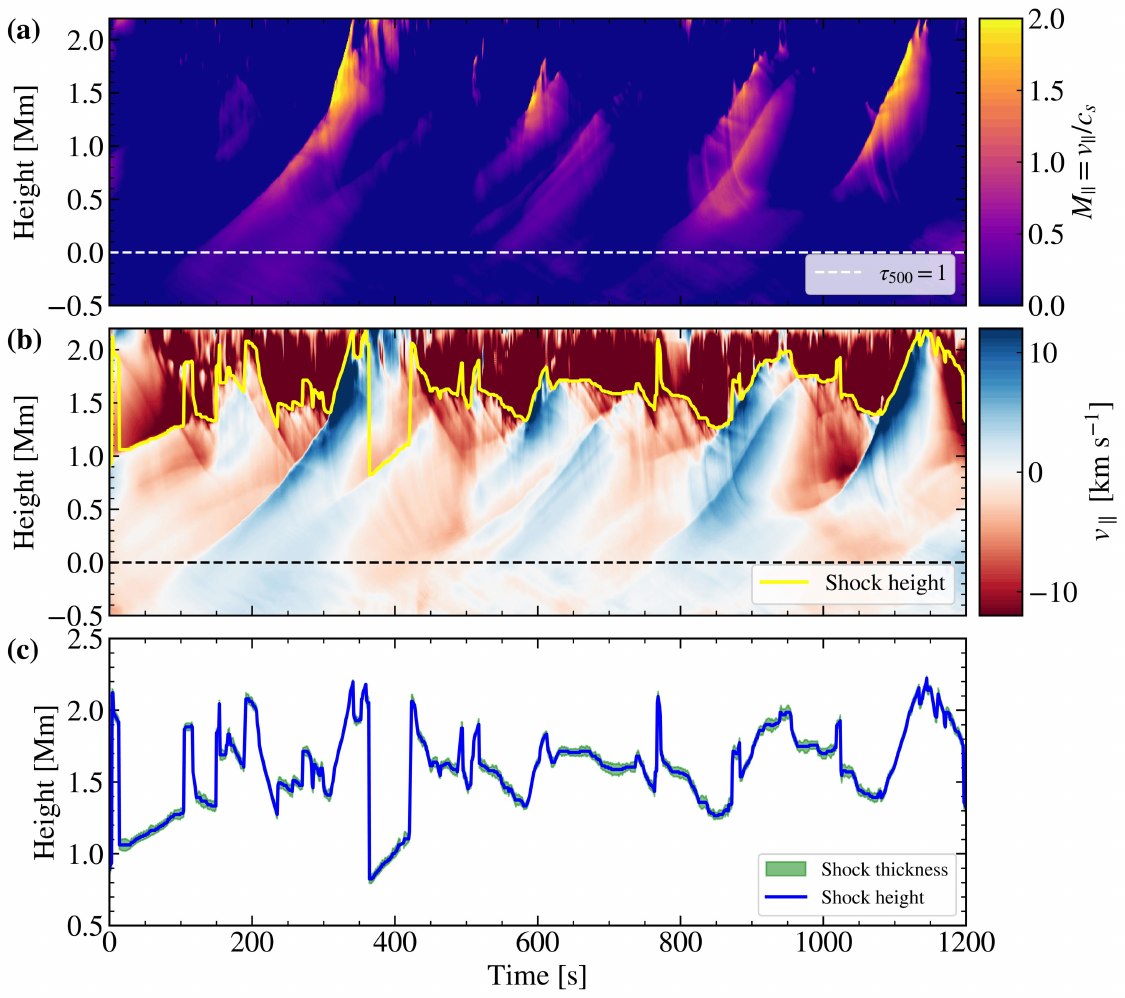}
\end{center}
\caption{(a) Height--time map of the Mach number computed from the field-aligned velocity and local sound speed, highlighting regions where the upflows becomes supersonic ($M > 1$). 
(b) Corresponding height--time map of the parallel velocity, with the shock locations overlaid in yellow, identified from the maximum gradient of the velocity along the field line. 
(c) Temporal evolution of the shock height (blue), together with the shock thickness (green).
}\label{fig:4}
\end{figure}

Figure~\ref{fig:4} summarizes the diagnostics used to identify and characterize shocks along the selected magnetic field line. The top panel shows the height--time evolution of the Mach number, computed using the field-aligned velocity and the local sound speed. Here, we plot a signed Mach number defined as $M_{\parallel} = v_{\parallel}/c_s$, and restrict the color scale to positive values in order to emphasize upward-propagating disturbances. As a result, regions with downflows ($M_{\parallel} < 0$) appear uniformly dark, allowing a clearer identification of supersonic upflows associated with shock propagation. We note that supersonic velocities are also present in downflowing plasma due to gravitational fallback; however, these do not necessarily correspond to shock structures and are therefore not emphasized in this visualization.
The middle panel shows the shock locations, highlighted in yellow, identified from the positions of the maximum gradient of the parallel velocity along the magnetic field line at each time step. These locations are overlaid on the height--time map of the parallel velocity, clearly tracing the upward propagation of supersonic disturbances. The detected shock fronts exhibit strong spatial and temporal correspondence with regions of enhanced Mach number in the upper panel, confirming that these steepened velocity features represent shock formation.
The bottom panel presents the temporal evolution of the shock thickness. The blue curve tracks the shock height as it propagates through the atmosphere, while the green shaded region shows the corresponding shock thickness. The thickness is estimated using the ratio of the velocity jump across the shock to the maximum velocity gradient, following Eq.~(2) of \citet{2024JCoPh.50512901I}.

It is important to note that the definitions of shock location and shock thickness adopted here should be interpreted with some caution. These diagnostics are traditionally formulated for idealized cases of one-dimensional, steady shocks propagating through a uniform medium with unidirectional flow. In contrast, the solar atmosphere considered here is strongly gravitationally stratified and highly dynamic. Radiative cooling in the chromosphere leads to the formation of dense plasma that can subsequently fall back under gravity, producing significant downflows coexisting with upward-propagating disturbances. As a result, the flow field is inherently multi-directional and time-dependent, and sharp gradients in velocity and density may arise not only from upward-propagating shocks but also from gravitationally-driven return flows, wave reflections, and shock-shock interactions. This complicates the identification of well-defined upstream and downstream states, and consequently introduces inherent ambiguity in the estimation of shock location and thickness based solely on instantaneous local gradients. For these reasons, the shock thickness derived in this work should be interpreted as an effective measure of the spatial extent of steepened compressive structures rather than a strict physical thickness of an ideal, steady shock. 
A more rigorous characterisation of these shock structures could in principle be achieved through comparison with generalised Rankine-Hugoniot jump conditions, which have been successfully applied in non-ideal plasma environments \citep{Snow2024}. Such an analysis, including a systematic investigation of how vortex dynamics influence shock width evolution, is deferred to a future study.

We note that the above analysis, based on a field line selected purely from magnetic field strength considerations, is entirely distinct from the vortex-specific analysis presented in the following section, where seed points are deliberately placed within identified vortex regions to systematically compare the thermodynamic and wave properties of vortex and non-vortex environments.
To examine the influence of photospheric vortex flows on the excitation, propagation, and dissipation of magnetoacoustic waves, we first identify regions within the selected domain that exhibit coherent rotational motions using the swirling strength computed from the three-dimensional velocity field. While the swirling strength diagnostic is, in principle, capable of capturing vortex structures without any filtering, in practice it is highly sensitive to small-scale velocity gradients and tends to identify numerous short-lived, fragmented vortical features embedded within larger coherent flows \citep{Yadav2020}. Such small-scale structures, although associated with locally enhanced rotation, do not necessarily represent the magnetically coherent vortex systems that persist across multiple atmospheric layers and are of primary interest in this study. As a result, the direct application of the method without filtering leads to significant spatial and temporal fragmentation, making it difficult to consistently track vortices and to associate them with magnetic field structures and wave dynamics.
To mitigate this issue, we have applied a horizontal Gaussian smoothing to the velocity field prior to computing the swirling strength. The smoothing length scale is chosen to vary with height, with minimal smoothing near the photosphere and progressively larger smoothing at higher layers, motivated by the expected expansion of coherent structures in a stratified atmosphere. This approach suppresses small-scale, short-lived fluctuations and facilitates the identification of larger, more persistent vortex structures that can be more reliably followed across heights and along magnetic field lines. We note, however, that such smoothing may modify the local vorticity amplitude, suppress fragmentation associated with vortex decay, and influence the inferred spatial extent and lifetime of vortices. Consequently, our analysis does not aim to provide precise quantitative estimates of vortex properties, but rather to identify representative, coherent vortex regions for examining their role in wave propagation and heating.

\begin{figure}[h!]
\begin{center}
\includegraphics[width=0.9\textwidth]{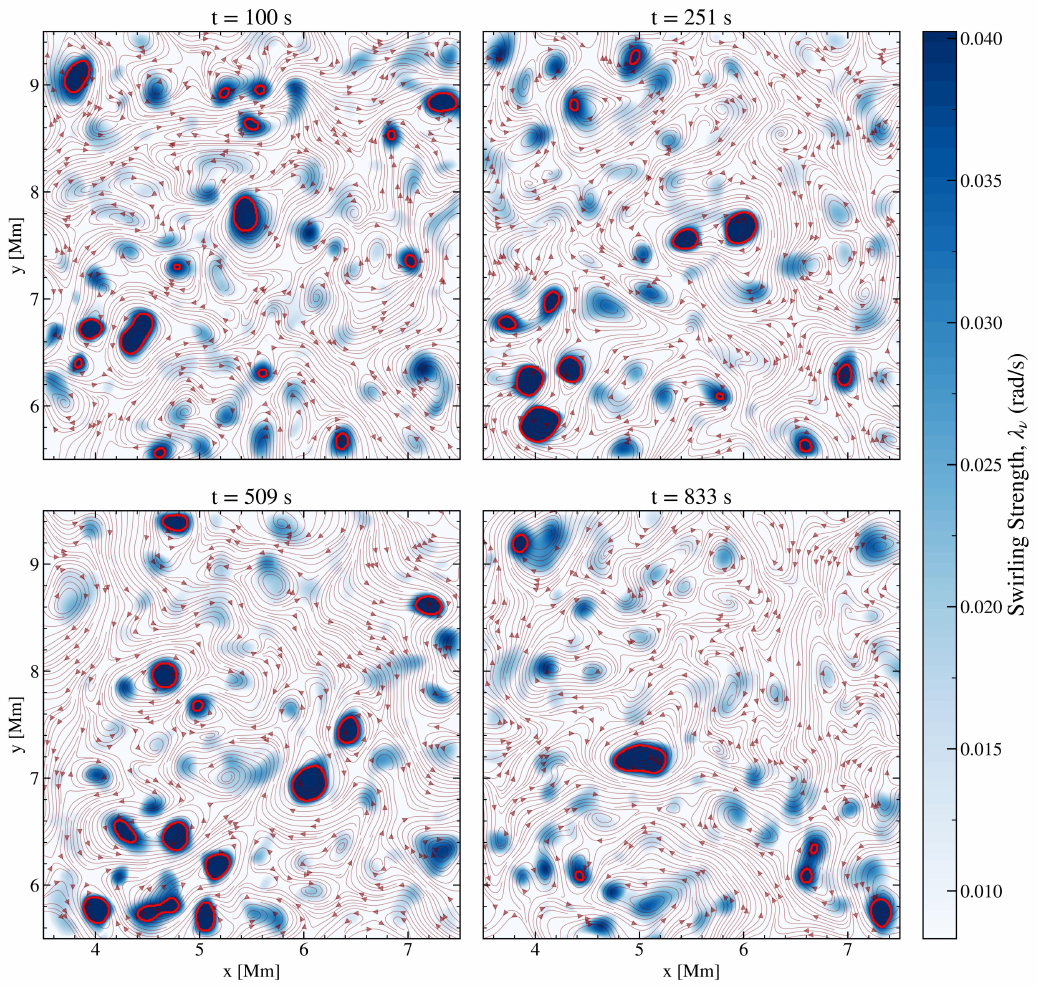}
\end{center}
\caption{Swirling strength map at z = 1 Mm at various time snapshots, overlaid with horizontal velocity vectors, highlighting coherent vortex structures. Red contours mark regions exceeding an additional threshold for strong rotation; these areas are used for magnetic field line tracing and for analyzing temperature and heating terms within vortex regions in further analysis}.\label{fig:5}
\end{figure}

Figure \ref{fig:5} presents maps of the swirling strength at a height of z = 1 Mm above the mean solar surface at different time snapshots, overlaid with the horizontal velocity vectors at the same layer. Clear signatures of rotational motion are evident, highlighting the presence of well-defined vortex flows. Regions exhibiting enhanced swirling strength correspond to locations where the plasma undergoes organized rotational motion, while areas with negligible swirling strength represent non-rotating regions devoid of vortex activity. 
In order to do a comparative study of thermal and dynamical evolution over voretx and non-vortex locations, an additional threshold ($\lambda_v > \mu + 2\sigma$, where $\mu$ and $\sigma$ are the mean and standard deviation of the swirling strength distribution  at each height) is applied to the swirling strength field. Regions exceeding this threshold are marked by red contours. Moreover, to select representative non-vortex regions with little or no rotation a threshold of ($\lambda_v < 0.001$\ rad/s) is applied.
% From these locations, magnetic field lines are traced in three dimensions, allowing us to follow the plasma and wave evolution along the natural magnetic pathways. 
We selected 150 seed points randomly distributed within identified swirling regions at z = 1 Mm layer (marked by red contours), from which magnetic field lines were traced in three dimensional simulation domain at each time snapshot.
% \textbf{To better illustrate the evolution of vortex structures across different atmospheric heights, we have  included a supplementary movie spanning from the solar surface to 2 Mm above the surface. This movie shows the temporal evolution of the flow field and vortex structures at multiple heights, providing a clearer picture of their vertical coherence and dynamics.}
In a magnetized plasma, swirling strength is not purely a localized scalar quantity but is dynamically coupled to the magnetic field and can extend along field lines as a consequence of MHD processes. As a result, regions of enhanced rotational motion tend to exhibit spatial coherence along magnetic structures.
Although magnetic field lines may undergo transverse displacements with height, our analysis shows that the traced field line continues to intersect regions of enhanced vorticity over a range of heights as can be seen in the movie associated with the Fig. \ref{fig:5}. Thus, the field line remains broadly associated with the vortex structure at all heights. The selected field line therefore provides a reasonable proxy for investigating the influence of vortex-associated dynamics on the thermodynamic and wave properties of the plasma as a function of height.

\begin{figure}[h!]
\begin{center}
\includegraphics[width=0.5\textwidth]{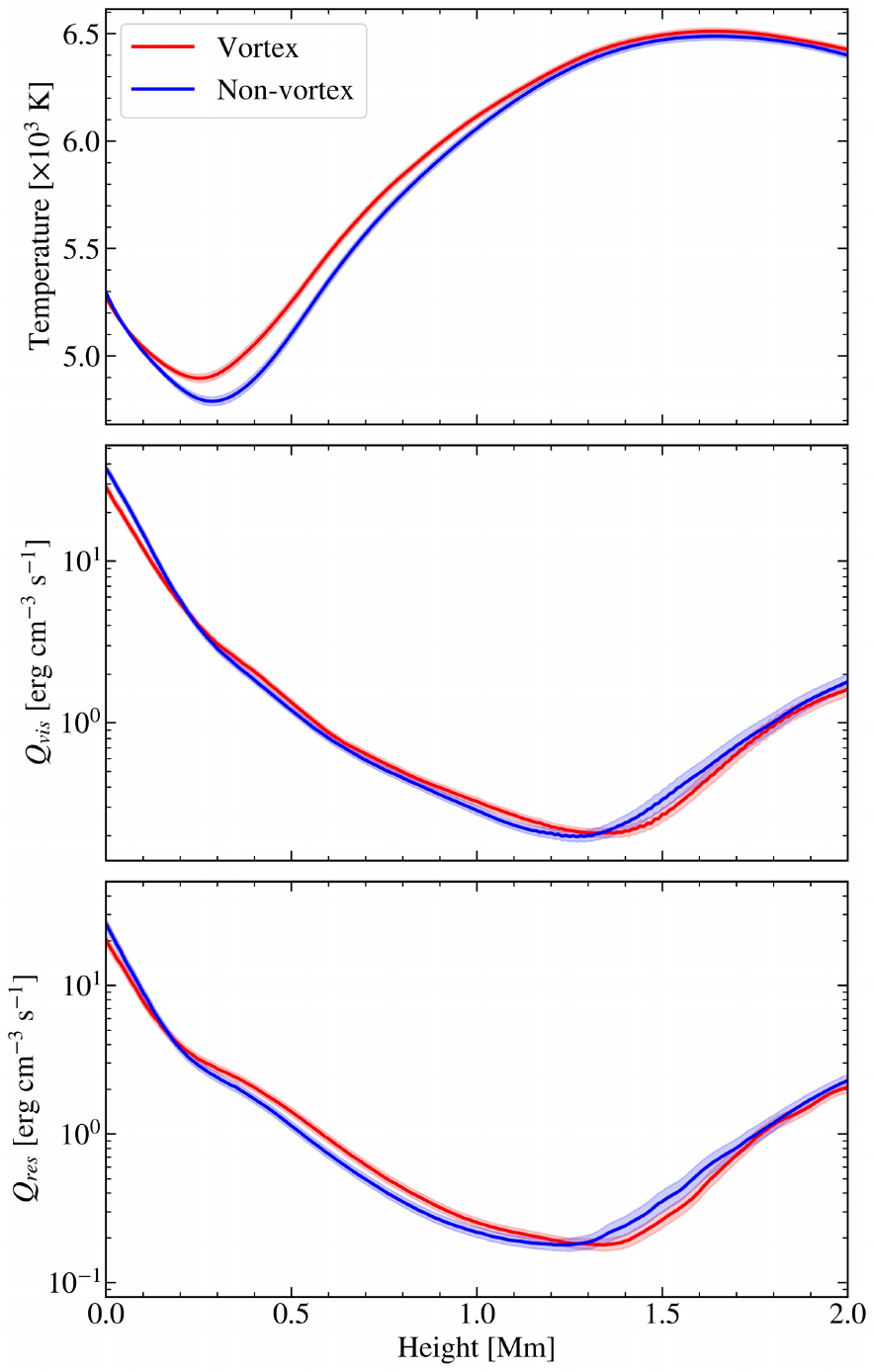}
\end{center}
\caption{Variation of (a) temperature, (b) viscous heating rate, and (c) Joule heating rate with height along magnetic field lines for vortex regions (red) and non-vortex  regions (blue). Solid lines represent the mean profiles averaged over all field lines and time snapshots, with shaded regions representing the standard error around the mean profiles.}\label{fig:6}
\end{figure}

Next, the temperature and heating rate profiles are obtained by averaging over all these field lines at a given snapshot, and subsequently over the temporal sequence as shown in Figure \ref{fig:6}. By examining the temperature distribution and heating-related quantities along these field lines, we directly compare the atmospheric response in vortex and non-vortex environments. This approach enables us to isolate the role of vortex-driven dynamics in modifying energy transport, wave propagation, and localized heating in the solar atmosphere.
Figure \ref{fig:6} compares the variation of temperature, viscous heating, and Joule heating as a function of height along magnetic field lines rooted in vortex and non-vortex regions. 
% In all three panels, the field line associated with the vortex region shows consistently higher values compared to the non-vortex case. 
The temperature profile exhibits a modest but systematic enhancement along the vortex-rooted field line, suggesting more efficient energy transport or deposition in these locations. 
Both viscous and Joule heating rates, show enhancement over vortices between 0.2-1.2 Mm height but lower values in the heights above. Although the explicit viscous and Joule heating rates do not show clear enhancements over vortices in comparison to non-vortex regions, the temperature along vortex-rooted field lines is systematically higher. This behaviour may be understood by noting that the temperature at these heights reflects the cumulative thermal history of the plasma, including heating deposited at lower heights that is subsequently advected upward along field-aligned structures, rather than only the instantaneous local dissipation. Furthermore, the shock fronts identified at heights of $\sim 1.3-2$ Mm contribute to compressive heating through entropy production, a mechanism that is not fully captured by the viscous and Joule dissipation terms alone. Together, these effects suggest that additional heating channels, beyond explicit resistive and viscous dissipation, are likely at play in sustaining the elevated temperatures within vortex regions at upper chromospheric heights.
We note that the enhancement in temperature and heating rate is not limited to a small subset of extreme cases, but is a consistent trend across the majority of field lines rooted in vortex regions. To quantify the spread in the data, we have include shaded regions representing the standard error around the mean profiles. This demonstrates that, while some variability exists, the overall enhancement is statistically robust and not dominated by a few outliers. 
This indicates that enhanced shock heating, arising from stronger wave amplitudes, sharper compressions, and increased shock dissipation, plays a dominant role in elevating the thermal state of vortex-associated plasma. The presence of vortex flows facilitates more efficient channeling and amplification of slow magnetoacoustic waves, leading to stronger nonlinear steepening and greater conversion of wave energy into heat.

\begin{figure}[h!]
\begin{center}
\includegraphics[width=0.8\textwidth]{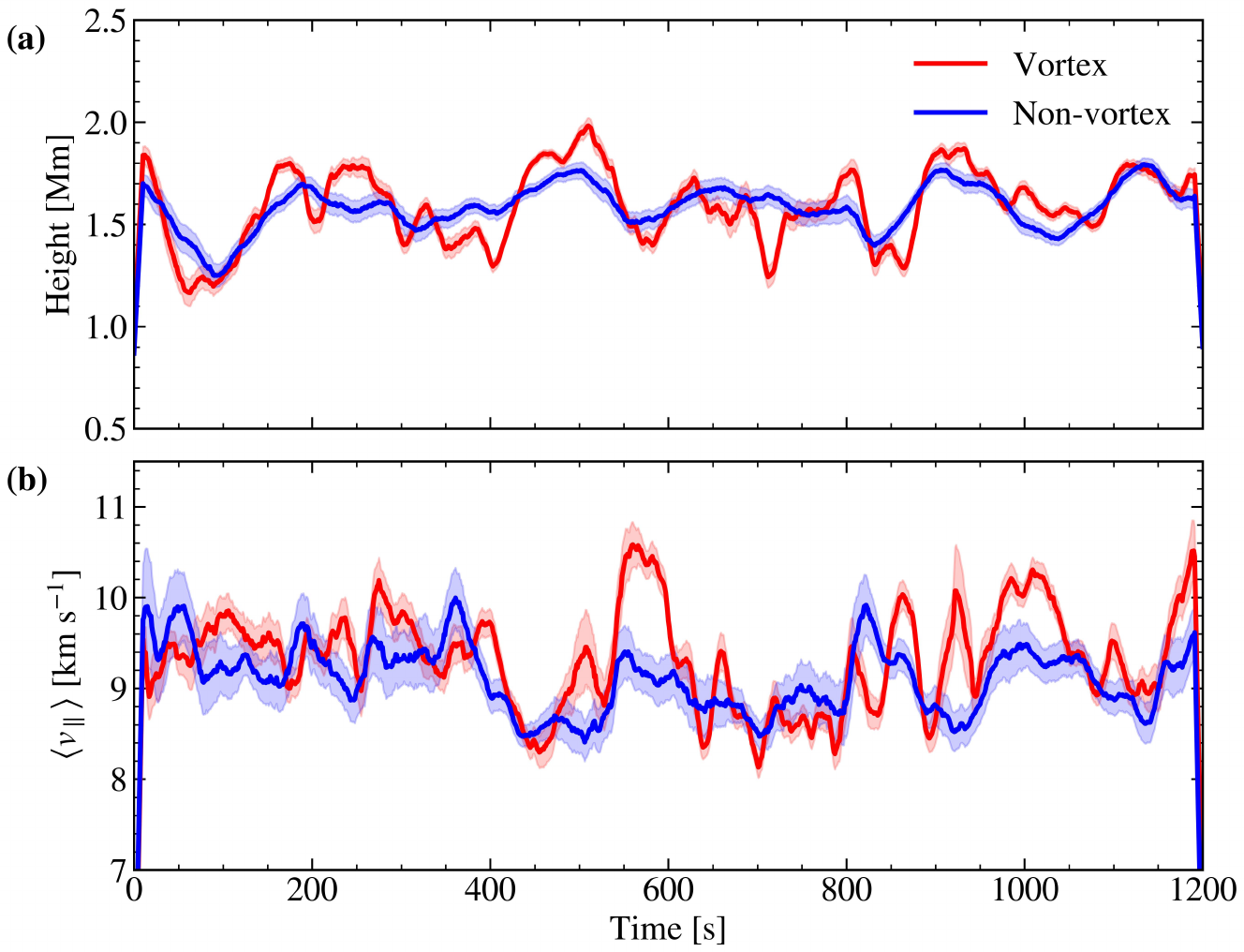}
\end{center}
\caption{a) Mean shock formation height and (b) mean $v_\parallel$ over time for vortex (red) and non-vortex (blue) regions. Solid lines show 20-point running averages with shaded bands indicating the standard error of the mean.}\label{fig:7}
\end{figure}

To investigate whether the presence of rotational flows influences the properties of acoustic shocks, we compare the mean shock formation height and mean parallel velocity of supersonic disturbances over time at vortex and non-vortex locations and display in figure \ref{fig:7}. For the shock formation height (panel a), we select 100 locations within strong vortex regions where the swirling strength exceeds a defined threshold (same as used earlier), and 100 locations in non-vortex regions where the swirling strength falls below a threshold (same as used earlier), ensuring that the two samples represent distinctly different rotational environments. For the parallel velocity analysis (panel b), we select 50 vortex locations where the swirling strength exceeds the threshold and which additionally exhibit supersonic upflows at a height of 1.5 Mm (in $\pm$ 200 km range), thus targeting vortex locations with presence of shocks. Figure~\ref{fig:7} shows the temporal evolution of the mean shock formation height (panel a) and the mean parallel velocity $\langle v_{\parallel} \rangle$ (panel b) for both populations, smoothed using a 20~s running average, with shaded bands representing the standard error of the mean. In panel (a) the two height profiles closely follow each other throughout the observation period, exhibiting no systematic offset or diverging trend, suggesting that rotational flows do not significantly alter the shock formation height. The average height at which disturbances become supersonic is approximately 1.5 Mm above the photosphere, consistent with mid-chromospheric heights typically associated with acoustic shock formation driven by p-mode oscillations. In panel (b), the vortex population shows somewhat higher parallel velocities at most times throughout the time series, suggesting that vortex-driven upflows may contribute to slightly enhanced supersonic velocities. This implies that while rotational flows do not fundamentally alter the shock formation height, they may play a role in amplifying the velocity of supersonic disturbances in the lower solar atmosphere.

\section{Summary}\label{summary}
In summary, we examined how slow magnetoacoustic waves propagate through the stratified solar atmosphere and how their evolution is influenced by photospheric vortex flows. 
By tracing magnetic field lines upward from the photosphere, we find clear signatures of wave propagation along the field. These waves originate near the photosphere with small amplitudes and behave like linear slow magnetoacoustic waves. As they move upward into the less dense chromosphere, their amplitudes increase due to the rapid drop in density. This amplification causes the waves to steepen and eventually form shocks. The passage of these shocks produces upward surges of dense plasma that follow parabolic trajectories, reaching heights of about 1–2 Mm before falling back under gravity. The periodicity of these motions, roughly five minutes, is consistent with solar p-mode oscillations. The associated changes in density, temperature, pressure, and velocity divergence all confirm the compressive and shock-like nature of these disturbances.

By tracking the density perturbations in space and time, we are able to identify the location and structure of the shock fronts. The shocks become sharper as they propagate upward, indicating nonlinear steepening. We also measure the shock thickness, which varies over time and reflects both the steepening process and the interaction of shocks with the surrounding plasma. In some cases, reflected shocks are observed, indicating complex interactions between upward-propagating disturbances and the stratified atmosphere.
We then investigated the role of vortex flows in modifying the atmospheric response. Vortex structures are clearly identified using the swirling strength diagnostic and are associated with coherent rotational plasma motions. Magnetic field lines rooted in vortex regions show systematically higher temperatures compared to those rooted in non-vortex regions. 
% \textcolor{blue}{Both viscous and Joule heating are slightly elevated in vortex regions, indicating relatively stronger velocity gradients and electrical currents}.
These results suggest that vortex flows contribute to localized energy dissipation and help enhance heating in the lower solar atmosphere.
Finally, we compared the shock formation height and the parallel velocity of supersonic disturbances between vortex and non-vortex regions. While the shock formation height evolves nearly identically for both populations with no systematic difference, the parallel velocities of supersonic disturbances are found to be somewhat higher over vortex locations with upflows at most times. This indicates that while vortex flows do not significantly alter the height at which shocks form, they may play a modest role in amplifying the velocity of supersonic disturbances. Together with the previously noted enhanced heating over vortex regions, this suggests that rotational flows contribute to energizing the local plasma environment, even if the shock generation mechanism itself operates largely independently of the rotational flow conditions.

Overall, our results show that slow magnetoacoustic waves naturally steepen into shocks as they propagate upward through the stratified atmosphere, transporting energy into the chromosphere. 
% At the same time, vortex flows provide an additional mechanism that enhances local heating and modifies the thermodynamic structure of the atmosphere. 
Vortex flows are known from previous studies to contribute to the stabilization and concentration of magnetic flux tubes. In such environments, the magnetic field remains more coherent and less susceptible to lateral diffusion, which can facilitate the efficient guidance of slow magnetoacoustic waves along the flux tube. In addition, rotational motions can introduce enhanced velocity shear, leading to stronger gradients in both velocity and thermodynamic quantities that may promote local dissipation and heating. However, the presence of elevated temperatures in vortex regions does not necessarily imply enhanced viscous or resistive dissipation alone. Other processes, such as adiabatic compression within converging flows, reduced radiative cooling due to local trapping of radiation, and the advection and confinement of hotter plasma, can also contribute significantly to the thermal structure of these regions. Furthermore, the mean height at which disturbances become supersonic is found to be approximately 1.5~Mm for both vortex and non-vortex regions, with no systematic difference in shock formation height, while supersonic disturbances over vortex locations with upflows exhibit somewhat higher parallel velocities, suggesting a modest dynamical imprint of rotational flows on shock amplitudes. We emphasize that these interpretations are currently speculative and based on physical reasoning supported by existing literature, and a more detailed and quantitative investigation is required to fully establish the dominant mechanisms responsible for the enhanced heating observed in vortex regions. The combined effects of wave propagation and vortex-driven dynamics likely play an important role in energy transport and heating in the lower solar atmosphere.

\section*{Conflict of Interest Statement}
%All financial, commercial or other relationships that might be perceived by the academic community as representing a potential conflict of interest must be disclosed. If no such relationship exists, authors will be asked to confirm the following statement: 

The authors declare that the research was conducted in the absence of any commercial or financial relationships that could be construed as a potential conflict of interest.

\section*{Author Contributions}

% The Author Contributions section is mandatory for all articles, including articles by sole authors. If an appropriate statement is not provided on submission, a standard one will be inserted during the production process. The Author Contributions statement must describe the contributions of individual authors referred to by their initials and, in doing so, all authors agree to be accountable for the content of the work. Please see  \href{https://www.frontiersin.org/about/policies-and-publication-ethics#AuthorshipAuthorResponsibilities}{here} for full authorship criteria.
Nitin Yadav conceived and supervised the study, developed the research framework, guided the methodology, and wrote the manuscript. Apanba Khuman performed all numerical analysis, including field-line tracking, shock calculations, and visualization of results. Both authors contributed to the interpretation of results, participated in scientific discussions, and reviewed and approved the final version of the paper.

\section*{Funding}
N.Y. received research funding from the DST INSPIRE Faculty Grant (IF21-PH-268), CSIR-ASPIRE grant (03WS[57291]/2023) and the SERB MATRICS grant (MTR/2023/001332).
A.K. received support from CSIR-ASPIRE grant (03WS[57291]/2023).

\section*{Acknowledgments}
Authors acknowledge the support from the DST, CSIR and ANRF.

% \section*{Supplemental Data}
%  \href{http://home.frontiersin.org/about/author-guidelines#SupplementaryMaterial}{Supplementary Material} should be uploaded separately on submission, if there are Supplementary Figures, please include the caption in the same file as the figure. LaTeX Supplementary Material templates can be found in the Frontiers LaTeX folder.

\section*{Data Availability Statement}
Data sets generated during the current study are available from the corresponding author on a reasonable request. 

% Please see the availability of data guidelines for more information, at https://www.frontiersin.org/about/author-guidelines#AvailabilityofData

\bibliographystyle{Frontiers-Harvard} %  Many Frontiers journals use the Harvard referencing system (Author-date), to find the style and resources for the journal you are submitting to: https://zendesk.frontiersin.org/hc/en-us/articles/360017860337-Frontiers-Reference-Styles-by-Journal. For Humanities and Social Sciences articles please include page numbers in the in-text citations 
\bibliography{test}

%%% Make sure to upload the bib file along with the tex file and PDF
%%% Please see the test.bib file for some examples of references

\section*{Figure captions}

%%% Please be aware that for original research articles we only permit a combined number of 15 figures and tables, one figure with multiple subfigures will count as only one figure.
%%% Use this if adding the figures directly in the mansucript, if so, please remember to also upload the files when submitting your article
%%% There is no need for adding the file termination, as long as you indicate where the file is saved. In the examples below the files (logo1.eps and logos.eps) are in the Frontiers LaTeX folder
%%% If using *.tif files convert them to .jpg or .png
%%%  NB logo1.eps is required in the path in order to correctly compile front page header %%%


% \begin{figure}[h!]
% \begin{center}
% \includegraphics[width=0.9\textwidth]{Delta_logrho_and_shock_thickness.eps}
% \end{center}
% \caption{Shock-height.}\label{fig:4}
% \end{figure}

% \begin{figure}[h!]
% \begin{center}
% \includegraphics[width=0.9\textwidth]{shock_thick.png}
% \end{center}
% \caption{Vertical variation of the mass density perturbation at one instant. The shock location (marked by red filled circle) is identified from the maximum vertical gradient of the density perturbation at that instant. The green and yellow circles mark the positions of the local maximum and minimum, respectively, that bracket the steepened profile.}\label{fig:5}
% \end{figure}

% \begin{figure}[h!]
% \begin{center}
% \includegraphics[width=0.9\textwidth]{swirling_strength_map.png}
% \end{center}
% \caption{Swirling strength map at z = 1 Mm, overlaid with horizontal velocity vectors, highlighting coherent vortex structures. Red contours mark regions exceeding an additional threshold for strong rotation; these areas are used for magnetic field line tracing and for analyzing temperature and heating terms within vortex regions. }\label{fig:6}
% \end{figure}

% \begin{figure}[h!]
% \begin{center}
% \includegraphics[width=0.9\textwidth]{neDelta_logrho_and_shock_thickness.eps}
% \end{center}
% \caption{Shock-height.}\label{fig:6}
% \end{figure}
%%% If you don't add the figures in the LaTeX files, please upload them when submitting the article.
%%% Frontiers will add the figures at the end of the provisional pdf automatically
%%% The use of LaTeX coding to draw Diagrams/Figures/Structures should be avoided. They should be external callouts including graphics.

\end{document}